\documentclass[twocolumn,superscriptaddress,amssymb,aps,pra]{revtex4-2}
\usepackage{amsmath,amsfonts,amssymb,amsthm,graphics,graphicx,epsfig,bbm}
\usepackage[colorlinks=true,citecolor=blue,linkcolor=blue,urlcolor=blue]{hyperref}
\usepackage[usenames]{color}

\usepackage{dsfont}
\usepackage{txfonts}
\usepackage{amsfonts}
\usepackage{mathrsfs}
\usepackage{physics}
\usepackage{subfigure}
\usepackage{color}
\usepackage{epstopdf}
\usepackage{bbold}
\usepackage{mathtools}
\usepackage{multirow}

\begin{document}
%%%%%%%%%%%%%%%%%%%%%%%%%%%%%%%%%%%%%%%%%%%%%%%%%%%%%%%%%%%%%%%%%%%

\title{Decoherences-protected implementation of quantum gates}

\author{Chunfeng Wu}
\email{chunfeng\_wu@sutd.edu.sg}
\affiliation{Science, Mathematics and Technology, Singapore University of Technology and Design, 8 Somapah Road, Singapore 487372, Singapore}

\author{Chunfang Sun}
\affiliation{Center for Quantum Sciences and School of Physics, Northeast Normal University, Changchun 130024, China}

\author{Jing-Ling Chen}
\affiliation{Theoretical Physics Division, Chern Institute of Mathematics, Nankai University, Tianjin 300071, China}

\author{X. X. Yi}
\affiliation{Center for Quantum Sciences and School of Physics, Northeast Normal University, Changchun 130024, China}

\begin{abstract}
We present a scheme to implement a universal set of quantum gates based on achievable interactions, and the gates can be protected against decoherences through dynamical-decoupling approach without encoding. By properly designing system evolutions, the desired system interactions commute with the elements forming dynamical decoupling pulses. Thus, the effect of decoherences can be eliminated by repeatedly applying the pulses, without noticeably affecting the system evolutions governed by the desired system interactions given small enough time interval between pulses. Moreover, due to the commutation between the elements forming the pulses and the desired system interactions, our scheme is resistant to different types of decoherences, and so not limited to specific decoherences. Our scheme also works well in the case that the desired system interactions cannot be achieved ideally due to imperfect control of system parameters, through the action of dynamical-decoupling pulses.
\end{abstract}

\date{\today}

\maketitle
\section{Introduction}
Realizing noise-resistant quantum operations at the level of physical qubits is of essential importance for further development of quantum computation. This is one of the main challenges faced by researchers to achieve fault-tolerant quantum computation, even though quantum error-correction codes (QECCs) have been extensively explored for actively detecting and correcting errors \cite{shor96,aharnov97,Kitaev97,Zurek98,NC,knill05,ion1,Cleland12,Meter12,Hollenberg11,Jones13,Brown21,Steane96,Shor95,Ruskai00,Ouyang14}. On the negative side, the success of QECCs replies on the requirements of physical-qubit resource and small enough errors \cite{Kitaev97,Zurek98,Hollenberg11,Jones13}. Because of the requirements, fault-tolerant quantum computer is beyond our reach with state-of-the-art technologies, still awaiting future developments in both theories and experiments.

Besides the QECCs, dynamical decoupling (DD) is also an active approach to fight against decoherences due to the interaction between system and environment \cite{Lidar14,dd1,dd2,dd3}. Specifically, DD fights against errors by using external pulse sequences to eliminate unwanted couplings, without observably causing impacts on the system dependent on the time interval between the pulses. Compared with QECCs, DD approach does not require a large number of physical qubits for encoding or measurement qubits for detecting errors, but needs a relatively acceptable number of control pulses which are achievable in different quantum systems \cite{Lyon06,Bollinger09,Bollinger092,Tombesi09,Davidson10}.

DD makes it possible to decouple physical qubits from the environment with comparably modest resource, ensuring the protection of the coherence of the qubits. However, applying the DD technique to the implementation of quantum gates cannot be achieved with ease since the DD pulses may aimlessly remove both the interaction between the system and environment, and the desired system interactions for executing quantum gates. How to make the DD technique compatible with quantum operations is the key task in achieving decoherences-protected quantum gates. The issue can be resolved through properly proposed system interactions \cite{Wun13}, or by encoding \cite{Viola1,Zanardi5,PRL89047901,PhysRevLett.100.160506,Gyure10} or by specially designed DD pulses \cite{Nature.484.82,NatCom13,NewJPhys.17.043008} to average out undesired terms and keep the terms required for implementing quantum gates. The general idea in the schemes is based on the harmony of gate Hamiltonians and DD pulses, for ensuring excellent performance of different gates in the presence of noises.

In this work, we present a scheme for executing decoherences-protected quantum gates formulated on properly designed system interactions, without encoding or specific DD pulses. Both single-qubit and two-qubit gates in our scheme can be actively preserved by different DD pulsese. Our scheme is based on two types of qubit-qubit interactions, and the desired interactions are achievable in superconducting systems with controllable system parameters  \cite{PRL89197902,PRL91057003,PRB81014505}. Two types of auxiliary qubits are needed for realizing a universal set of quantum gates. The use of auxiliary qubits preserves single-qubit gate operations from the disturbance caused by DD pulses. Meanwhile, the auxiliary qubits remain in their initial states after the evolution of system, and their coherence will be protected by the DD pulses as well. Neither measurement of auxiliary qubits nor replacement of used auxiliary qubits with fresh qubits is required. The desired system interactions commute with the elementary operations forming DD pulses. Thus, the effect of decoherences on both data qubits and auxiliary qubits can be eliminated by repeatedly applying the pulses, without evidently affecting quantum gate operations given small enough interval between the DD pulses. Moreover, due to the commutation between the decoupling elementary operations and the desired system interactions, our scheme is resistant to different types of decoherences and any DD approach constructed from the decoupling elementary operations is workable. Our scheme performs excellently even with some residuals in free system Hamiltonian which may be caused by imperfect control of system parameters, due to its compatibility with DD approach. The merit eases the requirement of strictly adjusting system parameters in actual experiments. Another important merit possessed by our scheme is that we do not need to change to the interaction picture to get desired gate Hamiltonians and as a result, the quantum gates in our scheme are preserved from decoherences in every step. Or else, further exploration is required for the protection of the system at the step of changing it back.

\section{Decoherences-protected quantum gates}
Our scheme is based on two types of system interactions,
\begin{eqnarray}
H_1=\frac{\epsilon}{2} (\sigma_z^i +\sigma_z^j) +\frac{\Delta}{2} (\sigma_x^i + \sigma_x^j) + J_z\sigma_z^i \sigma_z^j,
\end{eqnarray}
and 
\begin{eqnarray}
H_2=\frac{\epsilon}{2} (\sigma_z^i +\sigma_z^j) +\frac{\Delta}{2} (\sigma_x^i + \sigma_x^j) + J_x\sigma_x^i \sigma_x^j,
\end{eqnarray}
where $\epsilon$ and $\Delta$ describe the energy of qubits, $J_z$ is the coupling strength in $z$ direction, and $J_x$ is the coupling strength in $x$ direction. The two types of interactions can be implemented in superconducting qubits with controllable parameters $\epsilon$, $\Delta$, $J_{x}$ and $J_z$ \cite{PRL89197902,PRL91057003,PRB81014505}.

To implement single-qubit rotations about $z$ axis, we employ $H_1$ as 
\begin{eqnarray}
\label{h1}
H_1^{1\& A}=\frac{\epsilon}{2} (\sigma_z^1 +\sigma_z^{A}) + \frac{\Delta}{2} (\sigma_x^1 + \sigma_x^A)+ J_z\sigma_z^1 \sigma_z^{A}.
\end{eqnarray}
The Hamiltonian describes the coupling between data qubit 1 and auxiliary qubit $A$. We adjust the values of $\epsilon$ and $\Delta$ such that $\epsilon, \Delta\ll J_z$ during the system evolution. The evolution operator according to $H_1^{1\& A}$ is then of the form, $e^{-i\theta_1 \sigma_z^1 \sigma_z^{A}}$ with $\theta_1=J_z t$, where t is the evolution time. If initially the auxiliary qubit A is in its ground state $\ket{1}_A$, we get single-qubit rotations about $z$ axis on data qubit 1 as $U_1=e^{i\theta_1 \sigma_z^1}$.

Single-qubit rotations about $x$ axis on data qubit 1 can be achieved by utilizing another auxiliary qubit B from the evolution governed by $H_2$ of the following form,
\begin{eqnarray}
\label{h2}
H_2^{1\& B}=\frac{\epsilon}{2} (\sigma_z^1 +\sigma_z^{B})+\frac{\Delta}{2} (\sigma_x^1 + \sigma_x^B) + J_x\sigma_x^1 \sigma_x^B,
\end{eqnarray}
which represents the interation connecting data qubit 1 to auxiliary qubit $B$. 
Similarly we make the values of $\epsilon$ and $\Delta$ small enough to neglect the terms according to the condition $\epsilon, \Delta\ll J_x$. The resultant evolution operator  is $e^{-i\theta_2 \sigma_x^1 \sigma_x^{B}}$ with $\theta_2=J_x t$. If we set the auxiliary qubit B to be in its initial state $\ket{-}_B=\frac{1}{\sqrt{2}}(\ket{0}_B-\ket{1}_B)$, we obtain single-qubit rotations about $x$ axis on data qubit 1 and that is $U_2=e^{i\theta_2 \sigma_x^1}$. 

\begin{figure}[b!]
\centering
\includegraphics[width=3in]{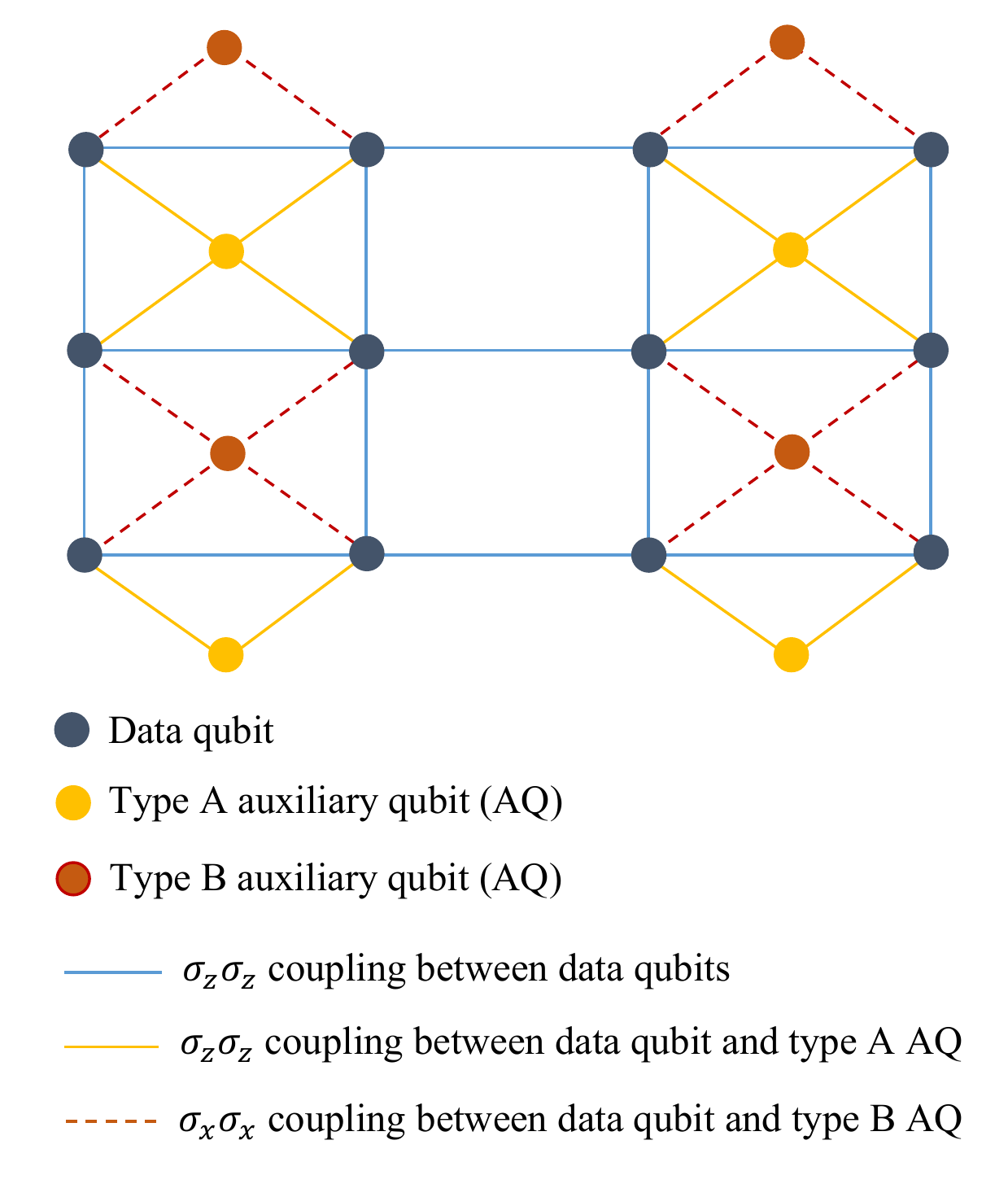}
\caption{The illustration of the couplings among data qubits and the two types of auxiliary qubits. Each data qubit is connected to one type A auxiliary qubit and one type B auxiliary qubit, and as results, arbitrary single-qubit gate can be achieved on each data qubit. Two neighbor data qubits are connected to each other in the form of $\sigma_z\sigma_z$ interaction or $\sigma_x\sigma_x$ interaction (this interaction is not demonstrated in the figure), in order to execute two-qubit gates on them.} \label{fig1}
\end{figure}

Two-qubit quantum gates can also be realized on the basis of $H_1$ expressing the interaction between two data qubits,
\begin{eqnarray}
\label{h3}
H_1^{1\& 2}=\frac{\epsilon}{2} (\sigma_z^1 +\sigma_z^{2})+\frac{\Delta}{2} (\sigma_x^1 + \sigma_x^2) + J_z\sigma_z^1 \sigma_z^{2}.
\end{eqnarray}
Again, we control the parameters $\epsilon$ and $\Delta$ properly to get an evolution operator $U_3=e^{-i\theta_3 \sigma_z^1 \sigma_z^{2}}$ with $\theta_3=J_z t$, without using any auxiliary qubit. The $U_3$ is locally equivalent to control-Z gate, namely $U_{\rm CZ}=e^{-i\frac{\pi}{4}}(S\otimes S)U_3|_{\theta_3=-\pi/4}$ with $S=e^{i\frac{\pi}{4}}U_1|_{\theta_1=-\pi/4}$. Moreover, the Hadamard gate can be achieved through the combination of the rotations about $x$ and $z$ axes in sequence, $H=-iU_2|_{\theta_2=\pi/2}U_1|_{\theta_1=-\pi/4}U_2|_{\theta_2=-\pi/4}U_1|_{\theta_1=\pi/4}$. The control-NOT gate is hence executable given the control-Z gate and the Hadamard gate, $U_{\rm CNOT}=(I_2\otimes H)U_{\rm CZ}(I_2\otimes H)$ (where $I_2$ is the $2 \times 2$ identity matrix). 

Moreover, another type of two-qubit quantum gates are achievable according to $H_2$ with the below coupling between qubits 1 and 2,
\begin{eqnarray}
\label{h4}
H_2^{1\& 2}=\frac{\epsilon}{2} (\sigma_z^1 +\sigma_z^{2})+\frac{\Delta}{2} (\sigma_x^1 + \sigma_x^2) + J_x\sigma_x^1 \sigma_x^{2}.
\end{eqnarray}
In the case that $\epsilon$ and $\Delta$ are adjusted to be small enough, we obtain an evolution opertor $U_4=e^{-i\theta_4 \sigma_x^1 \sigma_x^{2}}$ with $\theta_4=J_x t$. Therefore, we can implement a universal set of quantum gates on data qubits, including both Clifford and non-Clifford gates.

The reason that we employ auxiliary qubits is to make the DD approach compatible with the universal set of quantum gates. In other words, the auxiliary qubits are required to ensure that the desired coupling terms for implementing single-qubit gates will not be cancelled out by the DD pulses. Given the auxiliary qubits, the system evolution to realize single-qubit gates on the data qubit is governed either by Hamiltonian (\ref{h1}) or (\ref{h2}). The two types of Hamiltonians approximately commute with the decoupling group $\mathcal{G}=\{\openone^{\otimes N},\sigma_{x}^{\otimes N},\sigma_{y}^{\otimes N},\sigma_{z}^{\otimes N}\}$ \cite{Viola1,Zanardi5}, when $\epsilon$ and $\Delta$ are small enough. Even in the case that the two parameters cannot be effectively neglected, our scheme is still workable with the DD techniques. This is because the unwanted terms with the two parameters can be diligently eliminated during the system evolution with the application of DD pulses, while the desired terms for quantum gates commute with the above decoupling group. Similar reasoning goes to two-qubit quantum gates which are based on Hamiltonian (\ref{h3}). Therefore, desired coupling terms in Hamiltonians (\ref{h1},\ref{h2},\ref{h3}) will not be eliminated by the DD pulses and meanwhile undesired terms or decoherences will be effectively removed, leading to decoherences-protected quantum gates. 

From the above explanations, it is easy to see that our quantum gates are achieved in the Schr\"{o}dinger picture and we only control experimentally accessible system parameters to engineer the quantum gates. We do not need to change to the interaction picture to get desired gate Hamiltonians. This merit is important to protect quantum gates as the additional step of evolution to change the system back to the Schr\"{o}dinger picture is not required. Otherwise, we need to additionally explore how to protect the quantum gates when we change the system back to the Schr\"{o}dinger picture in order to fully preserve the quantum gates from noises.

In the literature, different methods of encoding logical qubits have been proposed to make the DD techniques adaptable with implementing quantum gates. Minimally, two physical qubits are required for encoding one logical qubit \cite{PRL89047901}. While in our scheme, the required computational resource can be reduced by removing encoding, but relying on auxiliary qubits. This is because one auxiliary qubit can be shared using by different data qubits, as illustrated in Fig. \ref{fig1}. There are two types of auxiliary qubits required in our scheme. We need to ensure each data qubit is connected to both types of auxiliary qubits, in order to achieve arbitrary decoherences-protected single-qubit gate on any data qubit. As shown in Fig. \ref{fig1}, 8 auxiliary qubits are needed for a quantum network with 12 data qubits, and hence totally 20 physical qubits are needed. With the encoding protocol \cite{PRL89047901}, a set of 24 physical qubits are required to achieve a quantum network with 12 logical qubits. Moreover, without the need of encoding, the complexity of measuring data qubits can also be mitigated.

\section{Numerical results}
In this section, we explore the performance of our decoherences-protected quantum gates by randomly choosing 100 initial states to find average fidelities, where $\ket{\psi}_0=\frac{1}{\sqrt{|a_0|^2+|a_1|^2}}\big(a_0\ket{0}_1+a_1\ket{1}_1\big)\ket{1}_A$ for $U_1$, $\ket{\psi}_0=\frac{1}{\sqrt{|a_0|^2+|a_1|^2}}\big(a_0\ket{0}_1+a_1\ket{1}_1\big)\ket{-}_B$ for $U_2$, and $\ket{\psi}_0=\frac{1}{\sqrt{\sum |a_{ij}|^2}}\sum_{i,j=0,1}a_{ij}\ket{i}_1\ket{j}_2$ for $U_3$ and $U_4$ with $a_{0,1}$ or $a_{ij}\in \mathbb{C}$. Specifically, we numerically calculate four types of gate fidelities, in the absence of decoherences, in the presence of decoherences but without applying DD pulses, in the presence of decoherences by applying periodic DD (PDD) pulses and concatenated DD (CDD) pulses, respectively. PDD can be expressed by ${\rm PDD}_{n_p}=(U_{\rm Blk})^{n_p}$, where $n_p$ is the number of times repeating $U_{\rm Blk}$ and $U_{\rm Blk}=\sigma_z^{\otimes N}U(T_0)\sigma_x^{\otimes N}U(T_0)\sigma_z^{\otimes N}U(T_0)\sigma_x^{\otimes N}U(T_0)$ formed by $U(T_0)$ (describes the evolution of the system with decoherences over $T_0$ with $T_0=\frac{t}{4n_p}$, where $t$ is desired evolution time) and the elements in the decoupling group \cite{Gyure10}. CDD is written as ${\rm CDD}_{n_c}=\sigma_z^{\otimes N}{\rm CDD}_{n_c-1}\sigma_x^{\otimes N}{\rm CDD}_{n_c-1}\sigma_z^{\otimes N}{\rm CDD}_{n_c-1}\sigma_x^{\otimes N}{\rm CDD}_{n_c-1}$ where $n_c$ is the number of times of repeating CDD base sequence and ${\rm CDD}_{1}=U_{\rm Blk}$ with $T_0=\frac{t}{4^{n_c}}$ for CDD \cite{Gyure10}.

In the presence of system-environment couplings, we assume the total Hamiltonian is written as $H_T=H_S+H_e$, where $H_S$ and $H_e$ are system Hamiltonian and the stochastic error term. In our calculations, $H_S$ is $H_1^{1\& A}$, $H_2^{1\& B}$,$H_1^{1\& 2}$, or $H_2^{1\& 2}$, and
\begin{eqnarray}
H_e=\sum_{k=1}^N(\delta_k^x\sigma_x^k+\delta_k^y\sigma_y^k+\delta_k^z\sigma_z^k),
\end{eqnarray}
where $N$ is the number of physical qubits and $\delta_k^{x,y,z}$ describe the strength of stochastic errors in different directions. The error model is selected according to Refs. \cite{diss1,diss2,diss3}, with $\delta_k^{x,y,z}$ randomly taken from a uniform distribution $[-J,J]$ where $J=J_x=J_z$. It is shown in Ref. \cite{diss2} that the dynamics of a system described by the Lindblad equation can be simulated by averaged dynamics governed by $H_T$ over stochastic errors. In our calculations, we choose 1000 sets of values of $\delta_k^{x,y,z}$ to model the stochastic process.

In the following, we study two cases of selecting different system parameters to show that quantum gates with desired fidelities are accomplishable in our scheme with applying DD pulses, no matter the term with $\epsilon$ or $\Delta$ can be neglected or not when DD techniques are employed. We choose $n_p=12$ and $n_c=3$ in the calculations, and round the fidelities up to four decimal places in Tables \ref{t1} and \ref{t2}.

{\it Case 1.} In this case, we choose the following parameters to investigate the performance of different gates, $\epsilon=2\pi\times 10$MHz, and $\Delta=0$, $J_x=J_z=2\pi\times 100$MHz. Here, we purposely set the value of $\epsilon$ to be small compared with $J_x=J_z$ and $\Delta$ to be zero. The numerical results are summarized in Table \ref{t1}. With small value of $\epsilon$ and zero value of $\Delta$, we can ignore the effect of the free Hamiltonians and obtain very good performance of the gates when decoherences are not taken into account. With the decoherences, the performance of the gates is largely affected by the decoherences without applying DD pulses. With DD pulses, we observe that both PDD and CDD preferably protect the quantum gates from the decoherences caused by the system-environment interaction in all $x$, $y$, and $z$ directions. The results are understandable since our desired interaction terms commute with the decoupling group. Any DD technique based on the decoupling group elements should be accomplishable in our scheme to eliminate the effect of decoherences. Dependent on the values of $\delta_k^{x,y,z}$, we can correspondingly adjust the number of the DD pulses to attain excellent gate fidelities in practical experiments. Moreover, we observe that the gate fidelities with the DD pulses are even better than those without decoherences. This is because that the non-zero free Hamiltonian term can be removed by the DD pulses too and hence the gate fidelities are further improved. 

\begin{table*}[ht]
\begin{tabular}{|c|c|c|c|c|}
\hline\hline
\multirow{3}{*}{Gate}  & \multicolumn{1}{c|}{Without Decoherences} & %
    \multicolumn{3}{c|}{With Decoherences}\\
\cline{2-5}
& Fidelity & Fidelity without DD & Fidelity with PDD ($n_p=12$) & Fidelity with CDD ($n_c=3$)   \\
\hline
$\sigma_x=-iU_2|_{\theta_2=\frac{\pi}{2}}$ & $ 0.9956$ & $ 0.3846$  & $ 0.9992$& $ 1$\\ \hline
$S=e^{i\frac{\pi}{4}}U_1|_{\theta_1=-\pi/4}$ & $ 0.9988$ &$ 0.5247$ & $ 0.9996$& $ 1$  \\ \hline
$T=e^{i\frac{\pi}{8}}U_1|_{\theta_1=-\pi/8}$ & $ 0.9997$ &$ 0.8235$ & $ 1$& $ 1$  \\ \hline
$U_3|_{\theta_3=-\pi/4}$ & $ 0.9973$  &$0.4562$ & $ 0.9997$& $ 1$\\ \hline
$U_4|_{\theta_4=-\pi/4}$ & $ 0.9978$  &$0.5002$ & $ 0.9999$& $1$\\ \hline
\hline
\end{tabular}
  \caption{The fidelities of a universal set of quantum gates in the absence (presence) of decoherences in Case 1. Further improvement in the fidelities  with decoherences can be achieved by increasing the number of times of repeating the base sequence of the DD pulses.} \label{t1}
\end{table*}

\begin{table*}[ht]
\begin{tabular}{|c|c|c|c|c|}
\hline\hline
\multirow{3}{*}{Gate}  & \multicolumn{1}{c|}{Without Decoherences} & %
    \multicolumn{3}{c|}{With Decoherences}\\
\cline{2-5}
& Fidelity & Fidelity without DD & Fidelity with PDD ($n_p=12$) & Fidelity with CDD ($n_c=3$) \\
\hline
$\sigma_x=-iU_2|_{\theta_2=\frac{\pi}{2}}$ & $0.6798$  & $ 0.3809$  & $ 0.9990$& $ 1$\\ \hline
$S=e^{i\frac{\pi}{4}}U_1|_{\theta_1=-\pi/4}$ & $ 0.8860$  &$ 0.5147$ & $ 0.9995$& $ 1$  \\ \hline
$T=e^{i\frac{\pi}{8}}U_1|_{\theta_1=-\pi/8}$ & $ 0.9707$  &$ 0.8053$ & $ 1$& $ 1$  \\ \hline
$U_3|_{\theta_3=-\pi/4}$ & $ 0.7524$ &$ 0.3882$ & $ 0.9996$& $ 1$\\ \hline
$U_4|_{\theta_4=-\pi/4}$ & $ 0.8033$  &$ 0.4458$ & $ 0.9999$& $ 1$ \\ \hline
\hline
\end{tabular}
  \caption{The fidelities of a universal set of quantum gates in the absence (presence) of decoherences in Case 2. Further improvement in the fidelities  with decoherences is attainable by increasing the number of times of repeating the base sequence of the DD pulses.} \label{t2}
\end{table*}

{\it Case 2.} We then revise $\epsilon=2\pi\times 100$MHz and keep the other parameters unchanged in the second case. Now the value of $\epsilon$ is not small enough to be ignored, comparable to $J_x=J_z$. We similarly find the average gate fidelities and the numerical results are summarized in Table \ref{t2}. It is understandable that the performance of the quantum gates is not as good as that in the first case when decoherences are not considered, due to the effect of the terms with parameter $\epsilon$. In this case, our scheme is still workable to achieve desirable gate fidelities with applying DD pulses since the undesired terms with the parameter $\epsilon$ can be cancelled out during the system evolution by the DD pulses. Repeat the base pulse sequence of PDD or CDD, it is shown clearly that the effect of the undesired terms is almost not visible given sufficient DD pulses. We do not even need to increase the number of times of repeating DD pulses as compared with Case 1. This is because that $\epsilon$ is of the same magnitude as the strength of the stochastic errors. The DD pulses can eliminate the effects due to the term with parameter $\epsilon$ (and similarly $\Delta$) and decoherences. Again, our scheme is compatible with different DD techniques, due to the commutation of the desired coupling terms and the decoupling group elements. Moreover, the gate fidelities can be improved by increasing the number of times of applying the DD pulses. With increasing value of $\epsilon$ or $\Delta$, more number of times of repeating base DD sequence will be required to eliminate the unwanted terms. The result is friendly and useful in practical experiments if there are some uncontrollable residuals even though $\epsilon$ and $\Delta$ are adjustable.

\section{Discussion and Conclusion}
We have explored the implementation of a universal set of quantum gates that can be protected against decoherences by the DD techniques, based on achievable interactions. The compatibility of implementing quantum gates and the DD techniques is on the basis of employing auxiliary qubits, rather than encoding. For single-qubit gates, we utilize auxiliary qubits and proper system interactions to execute decoherences-protected rotations about $x$ and $z$ axes, ensured by the fact that the desired system interactions commute with the elements forming DD pulses. For two-qubit gates, no auxiliary qubit is needed for implementing decoherences-protected non-trivial quantum gates. Because of the commutation between the elements forming DD pulses and the desired system interactions, the effect of different types of decoherences can be eliminated by repeatedly applying DD pulses, without manifestly affecting the system evolutions. Moreover, compared with the protocols with encoding, our scheme relaxes the need of qubit resource and reduce the complexity of measuring data qubits. In pratical experiments due to control capabilities, there may be some residuals in free system Hamiltonian besides the desired interactions. We illustrate that our scheme is not sensitive to the residual terms, with DD pulses applied properly. On another positive side, our scheme is based on the Hamiltonians in the Schr\"{o}dinger picture rather than in the interaction picture and thus, there is no need of additional step of evolution back to the Schr\"{o}dinger picture. This merit simplifies the process of protecting the quantum gates from the decoherences. Otherwise, we need to explore the protection of the quantum gates in the additional step of evolving back to the Schr\"{o}dinger picture as well, in order to achieve full protection of the quantum gates.

Given the decoherences-protected Clifford and non-Clifford gates in our scheme, arbitrary quantum operation can be protected actively against decoherences. Thus our scheme is of practical importance in resolving various quantum computation tasks. Here as an example, we would like to discuss the application of our protected quantum gates in the implentation of QECCs. As mentioned in Ref. \cite{PRA84012305}, the integration of the DD and QECCs leads to more powerful platform for preserving quantum information from errors due to the (dis)advantages of either method.

The first example to discuss is surface codes that are widely explored for achieving fault-tolerant quantum computation \cite{Cleland12,Meter12}. The syndrome measurements are crucial for surface codes since the implementation of surface codes depends on the syndrome measurements aided by desired quantum operations acting on computational qubits \cite{Meter12}. The syndrome measurements can be executed by acting Hadamard and CNOT gates in sequence on measurement and data qubits, together with certain measurements \cite{Cleland12}. Therefore, based on our protected quantum gates, it is possible to achieve protected implementation of surface codes.  

Besides the possible application in surface codes, our scheme is also beneficial for protecting the implementation of other types of QECCs. For most of the QECCs, the implementation of the codes conventionally relies on single/two-qubit quantum gates acting on physical qubits \cite{NC}. Specifically, twelve control-Z gates are needed for generating the Steane code, and five control-phase gates plus several single-qubit gates are employed in the encoding circuit of the five-qubit code \cite{Kwekl5}. Consider the number of quantum gates required, it is highly essential to preserve quantum information from decoherences during the multi-step system evolution. The decoherences-protected quantum gates presented in our scheme can enhance the success probability of implementing the QECCs. Furthermore, our protected quantum gates are also helpful in the initialization of the permutation-invariant QECCs, in the case that the implementation of the codes is mainly not based on single/two-qubit operations \cite{usnew}.

Our scheme is also helpful for executing quantum machine learning in physical systems. In Ref. \cite{ML}, the authors explored the creation of handwritten digits with high resolution on the basis of quantum circuits formed by different gates, showing the power of quantum computers in enhancing the performance of machine learning algorithm. The desired quantum gates in the quantum circuit are preservable from decoherences according to our scheme. Therefore, the quantum machine learning protocol can be protected by DD techniques. 

To conclude, our scheme plays favourable roles in boosting the success rate of performing different quantum-computation tasks in the presence of decoherences. The robust merit and the achievability of desired system interactions in superconducting systems signify the scheme as an important step forward towards the development of fault-tolerant quantum computation.

\vspace{5pt}
C.W. is supported by the National Research Foundation, Singapore under its QEP2.0 programme (NRF2021-QEP2-02-P03). C.S. is supported by Fundamental Research Funds for the Central Universities (Grant No. 2412019FZ040). J.L.C. is supported by National Natural Science Foundation of China (Grant No. 11875167 and 12075001). X.X.Y. is supported by National Natural Science Foundation of China (Grant No. 12175033).

\vspace{8pt}

\end{document}